\documentclass[a4paper,11pt]{article}
\pdfoutput=1 

\usepackage{jinstpub} 

\usepackage{lineno}

\title{\boldmath Performance of photosensors in a high-rate environment for gas Cherenkov detectors}

\author[a]{C. Peng,}
\author[a,1]{J. Xie,\note{Corresponding author.}}
\author[a]{S. Joosten,}
\author[a]{Z.-E. Meziani,}
\author[b]{M. Jones,}
\author[c]{E. Kaczanowicz,}
\author[c]{M. Rehfuss,}
\author[c]{N. Sparveris,}
\author[d]{M. Paolone,}
\author[e]{M. Foley,}
\author[e]{M. Minot}
\author[e]{and M. Popeckie}

\affiliation[a]{Argonne National Laboratory, 9700 S Cass Ave., Lemont, IL 60439, USA }
\affiliation[b]{Thomas Jefferson National Accelerator Facility, 12000 Jefferson Ave., Newport News, VA 23606, USA }
\affiliation[c]{Department of Physics, Temple University, Philadelphia, PA 19122, USA }
\affiliation[d]{Department of Physics, New Mexico State University, Las Cruces, NM 88003, USA }
\affiliation[e]{Incom, Inc., 294 Southbridge Rd., Charlton, MA 01507, USA }

\emailAdd{jxie@anl.gov}

\abstract{The solenoidal large intensity device (SoLID) at Jefferson Lab will push the boundaries of luminosity for a large-acceptance detector, which necessitates the use of a light-gas threshold Cherenkov counter for online event selection. Due to the high luminosity, the single-photon background rate in this counter can exceed 160~kHz/cm$^2$ at the photosensors. Therefore, it is essential to validate the high-rate limits of the planned photosensors and readout electronics in order to mitigate the risk of failure. We report on the design and an early set of studies carried out using a small telescopic Cherenkov device in a high-rate environment up to 60~kHz/cm$^2$, in Hall~C at Jefferson Lab. Commercially available multi-anode photomultipliers (MaPMT) and low-cost large-area picosecond photodetectors (LAPPD) were tested using the JLab FADC250 modules for readout. The test beam results show that the MaPMT array and the internal stripline LAPPD can detect and identify single-electron and pair-production events in high-rate environments. Due to its higher quantum efficiency, the MaPMT array provided a better separation between the single-electron and the pair-production events compared to the internal stripline LAPPD. A GEANT4 simulation confirms the experimental performance of our telescopic device.}


\arxivnumber{2011.11769} 

\begin{document}
\maketitle
\flushbottom

\section{Introduction}
\label{sec:intro}

The Continuous Electron Beam Accelerator Facility (CEBAF) in the 12 GeV era of Jefferson Lab is able to deliver a high-intensity continuous electron beam simultaneously to four experimental halls. 
Jefferson Lab employs its experimental halls in complementary pairs: Halls A and C combine very high beam currents --- upwards of 100~$\mu$A --- incident on extended targets with narrow-acceptance spectrometers, while Halls B and D combine a more moderate beam current with large-acceptance detectors.
The future experiments using the proposed SoLID detector will operate a large-acceptance detector in Hall A at high beam currents~\cite{1,2}.
This unprecedented luminosity requirement for a large-acceptance detector imposes new challenges on detector technology, readout and trigger design, and data acquisition.
The cornerstone SoLID experiments require operation at a luminosities between 10$^{37}$--10$^{39}$~cm$^{-2}$s$^{-1}$ \cite{2}. 
In these experiments, the SoLID high-threshold light-gas Cherenkov detector (LGC) will be crucial to identify electrons and reject pions for online event selection.

Traditional Cherenkov threshold counters, using a gas radiator (e.g., N$_2$, CO$_2$, CF$_4$, or C$_4$F$_{10}$), serve as particle-identification detectors capable of separating different particle species based the radiator choice and particle velocity.
For example, a light-gas radiator provides a cone of Cherenkov light for electrons passing through the medium while leaving no response for pions with energies below the radiator's energy threshold. The crucial challenge for gas Cherenkov detectors in high luminosity experiments is the exceedingly high rate of background photons. For the SoLID LGC, a typical signal event will have over 30 photons in a Cherenkov disk of approximately 10~cm diameter at the face of the photosensors. In contrast, the single-photon background rate incident on the photosensors, caused by different background sources (e.g., $\pi^0$ decay), is projected to exceed 4~MHz per $5 \times 5$~cm$^2$ area. High-granularity photosensors allow for this high-rate uncorrelated background to be separated from good Cherenkov rings at the trigger level. Furthermore, the SoLID magnetic field at the LGC's photosensor position measures about 200 gauss, making a traditional Cherenkov approach with larger 3- or 5-inch quartz-glass dynode-type photomultipliers impractical due to their magnetic field sensitivity. Multi-anode photomultiplier tubes (MaPMTs) \cite{3,4,5,6} and microchannel plate photomultipliers (MCP-PMTs) \cite{7,8,9,10} are good options for the LGC due to their high-granularity photosensors and relative radiation hardness. 
Compared to MaPMTs, MCP-PMTs have higher magnetic field resiliency, while MaPMTs require proper magnetic field shielding for applications in magnetic field environments.
Both sensors can be tiled due to their square shape, and are therefore able to cover large detection areas. 

Currently, 64-pixel MaPMTs are commercially available from Hamamatsu \cite{11}. They are sensitive to the spatial distribution of intensity across their surface and can be used to replace traditional tubular PMTs in many experimental devices \cite{12,13,14,15,add1,add2}. Compared to standard quartz-glass PMTs, MaPMTs have a square shape and high granularity, with a pixel pitch of $6 \times 6$ mm$^2$. 
A p-terphenyl wavelength shifter coating can enable a high photodetection efficiency at shorter wavelengths in the UV spectrum~\cite{16}.
One major drawback of MaPMTs is their relatively lower magnetic field tolerance, requiring the use of full-array mu-metal shielding and surrounding soft iron shielding in magnetic fields above 50 gauss to ensure optimal photodetection efficiencies~\cite{17}. 
The additional shielding not only complicates the detector design but also increases the final detector cost. 
MCP-PMTs are built upon an advanced technology that employs microchannel plates to replace conventional discrete dynodes for electron signal amplification \cite{10}. Commercial MCP-PMTs have been shown to have superior timing and position resolution, high granularity, and high magnetic field tolerance \cite{18,19}. These advantages make MCP-PMTs an ideal photosensor candidate: they enable a simplified detector design while operating at high efficiencies. 
However, the current generation of MCP-PMTs is significantly more expensive than contemporary MaPMTs, preventing their use for large-scale experiments due to cost concerns.
A novel, recently commercialized MCP-PMT, the Large Area Picosecond Photo-Detector (LAPPD)~\cite{20,21,22}, provides a promising cost-effective MCP-PMT for Cherenkov photon counting and imaging applications. The LAPPDs use low-cost micro-channel plates activated by applying resistive and secondary emissive layers on a large-area glass capillary substrate through the atomic layer deposition (ALD) technique. The cost of LAPPDs is expected to be at least an order of magnitude lower per active area than that of traditional MCP-PMTs \cite{23}. For a given sensor coverage area, this projected cost reaches a price range competitive to that of MaPMTs. Because of this, the LAPPD is a strong candidate to be used as photosensor for future Cherenkov counters. 

Due to their high granularity, a typical Cherenkov cone will illuminate multiple sensitive areas of MaPMTs and MCP-PMTs, generating spatially distributed signals. A simple coincidence requirement that at least two nearby channels fire can then drastically reduce the accidental single-photon background from random sources. 
Undoubtedly, this novel application of the MaPMTs and MCP-PMTs in a harsh environment with a very high single-photon background rate requires extensive testing to guide design decisions and provide input for realistic simulations. In this paper, we report on the design of a small Cherenkov telescope, along with the photosensor evaluation test beam exercise performed at the Jefferson Lab. We installed the small detector prototype for parasitic operation during an on-going fixed-target experiment in a realistic “open” high-rate environment in Hall C. We evaluated both commercially available MaPMTs and an LAPPD as photosensors. We observed clear Cherenkov signals originating from energetic electrons, using a calorimeter as main trigger. The performance of the photosensors in this high-rate/high-background environment will inform their use in future gas Cherenkov counters. 

\section{Small Cherenkov telescope design}
\label{sec:design}
\begin{figure}[htbp]
\centering 
\includegraphics[width=1.0\textwidth,trim=0 110 0 50,clip]{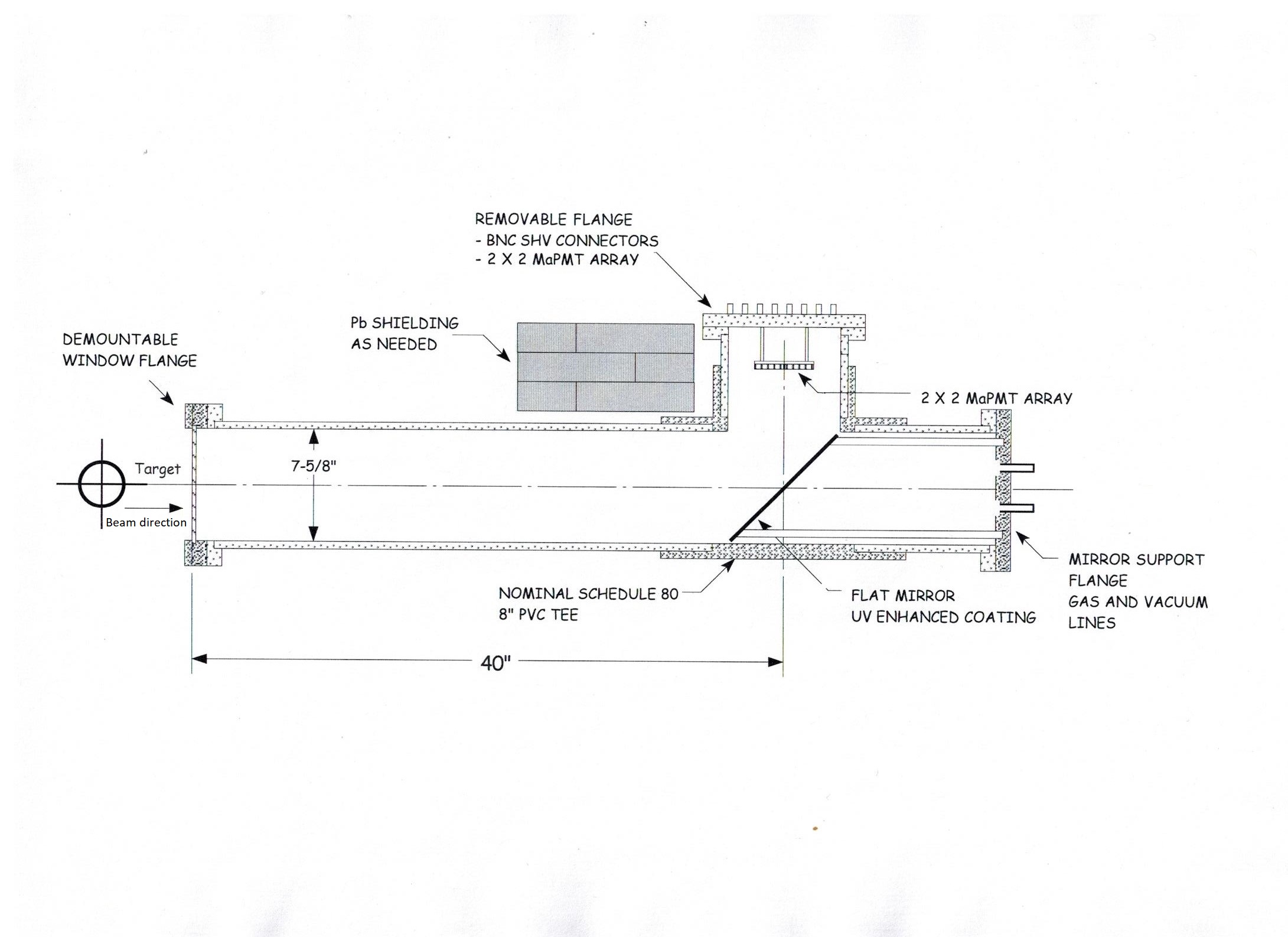}
\caption{\label{fig1} Schematic of the small telescopic Cherenkov detector.}
\end{figure}

We constructed a small Cherenkov telescope, shown in Figure~\ref{fig1}, to test MaPMTs and an LAPPD in a realistic high-rate/high-background environment. The device consists of four components: (1) the main detector tank, (2) the flat reflective mirror, (3) the gas pressure monitor system, and (4) the photosensor. 

The main detector tank was constructed out of standard schedule-80 PVC pipes, with an inner diameter of 7-5/8 inches. A tee-socket was used to connect the three individual PVC pipes into a telescope shape. The 40-inch-long pipe faced the target during the test beam exercise to allow scattered electrons to pass through the gas radiator and generate Cherenkov photons. The radiator was kept just above atmospheric pressure by an entrance and exit windows made out of aluminum of 75 $\mu$m thickness. A flat mirror\footnote[2]{Spectra Thin Films, Hauppauge, NY} coated with a thin aluminum film for enhanced ultraviolet (UV) light reflection was placed in the center of the tee-socket at an angle of $45^\circ$ with respect to the center of the main pipe. The mirror reflected the Cherenkov photons by $90^\circ$ to land on the surface of the photosensors, which were mounted on the side port of the tee-socket and protected from radiation damage by lead blocks. The interior of the entire system was coated with light-absorbing paper to minimize internal reflection. A 370 nm LED was installed near the entrance window for calibration and testing purposes. 

A gas system from the SANE Cherenkov counter \cite{24} was re-purposed and installed on the Cherenkov telescope to regulate the gas pressure slightly above atmospheric pressure. Operating at atmospheric pressure helped simplify the mechanical design and minimized the windows thicknesses. CO$_2$ gas, with a reflective index $n = 1.000448$ at atmospheric pressure and room temperature, was used as the radiator medium in this system. This resulted in a Cherenkov light cone with a critical angle $\theta_c \sim 1.7^\circ$ for electrons with momenta above 17 MeV/c. The Cherenkov tank was flushed in several cycles to ensure a gas purity above 99\%.

\begin{figure}[htbp]
\centering 
\includegraphics[width=1.0\textwidth,trim=0 0 0 0,clip]{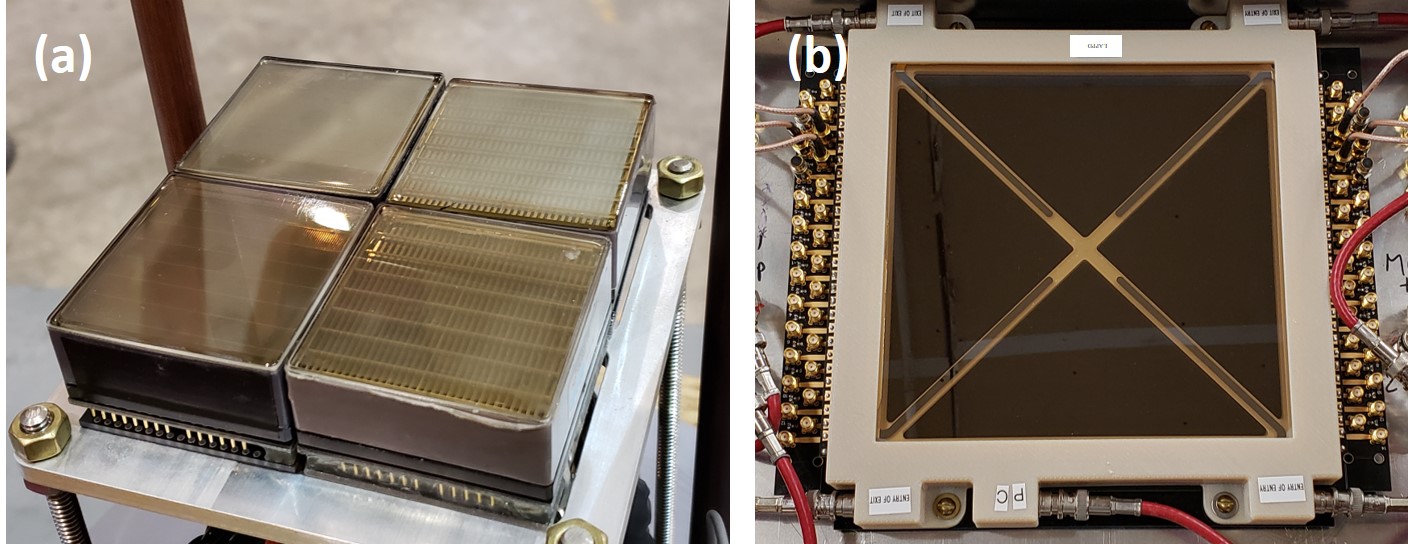}
\caption{\label{fig2} (a) $10 \times 10 $~cm$^2$ p-terphenyl coated MaPMT array, and (b) $20 \times 20$~cm$^2$ LAPPD.}
\end{figure}

An array of $2 \times 2$ Hamamatsu MaPMTs was tiled to provide a large photosensor area of roughly $10 \times 10$~cm$^2$. The array consisted of two H8500C-03 and two H12700A-03 MaPMTs. Each MaPMT comprises a square grid of $8 \times 8$ pixels with 5.80~mm pitch. The SoLID LGC does not require such fine-grained pixel pitch, hence we read out each MaPMT into quadrants of 16 pixels each, reducing the number of readout channels. This allows for cost savings in the readout electronics, and increases the number of photoelectrons per channel. 
In total, we had 16 channels for the $2 \times 2$ array of Hamamatsu MaPMTs (four channels per PMT). Figure~\ref{fig2}(a) shows an image of the $2 \times 2$ MaPMT array. All four MaPMTs were coated with a p-terphenyl wavelength shifter to enhance their response to UV photons~\cite{16}. A low-cost internal stripline LAPPD (LAPPD $\#41$) was provided by Incom, Inc. to evaluate its performance in a high-rate environment. It features an active area close to $20 \times 20$~cm$^2$ and has a stripline readout with 28 strips \cite{21}. Pixelized LAPPDs were still under development and not yet commercially available when our test beam exercise was conducted. The performance of a pixelized device will be characterized in a future measurement. Figure~\ref{fig2}(b) shows an image of the internal stripline LAPPD. Due to its large size, a dark adaptor box was designed and built to accommodate the LAPPD. The provided LAPPD was fabricated with a fused silica entrance window. Bialkali photocathode was coated on the inside surface of the entrance window, providing photon detection down to 220 nm. The geometry and performance parameters of the Hamamatsu MaPMTs, and Incom stripline LAPPD are listed in table~\ref{tab1}. 

\begin{table}[htbp]
\centering
\caption{\label{tab1} Geometry and performance parameters of photosensors to be investigated}
\smallskip
\begin{tabular}{|l|l|l|l|}
\hline
  & H8500C-03	& H12700A-03	& LAPPD \#41\\
\hline
PMT type &	MaPMT	& MaPMT &	MCP-PMT\\
Manufacturer & Hamamatsu & Hamamatsu & Incom\\
Device type &	12-stage dynode	& 10-stage dynode &	Gen-I LAPPD\\
Active area  &	5~$\times$~5~cm$^2$  &	5~$\times$~5~cm$^2$  &	20~$\times$~20~cm$^2$\\
Readout type  &	Pixel &	Pixel  &	Stripline\\
Entrance window material  &	UV glass  &	UV glass  &	Fused silica\\
Wavelength shifter coating  &	Yes  &	Yes  &	No\\
Response wavelength range  &	185 – 650 nm  &	185 – 650 nm  &	220 – 700 nm\\
Peak quantum efficiency (QE)  &	28\% (400 nm)  &	35\% (380 nm)  &	7.3\% (365 nm)\\
Operation high voltage  &  950 V  &	950 V  & 2550 V (MCP = 975 V)\\
Gain  &	106  &	106  &	4 × 106\\
\hline
\end{tabular}
\end{table}
	
\section{Experimental setup}
\label{sec:Exper}
	
\begin{figure}[htbp]
\centering 
\includegraphics[width=1.0\textwidth,trim=0 0 0 0,clip]{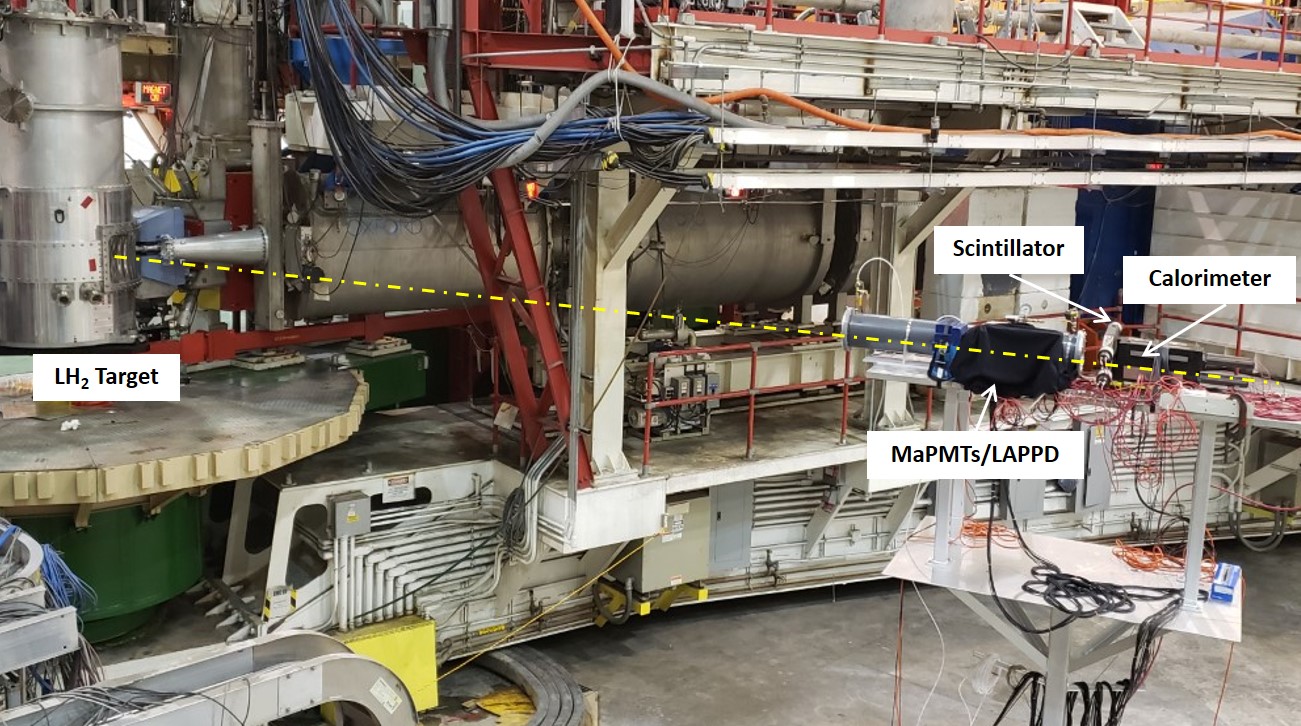}
\caption{\label{fig3} Cherenkov telescope installed on a platform in Hall C. The setup was lifted to the level of the beamline, 4.8 meters away from the liquid hydrogen target, and with the entrance window facing the target chamber. Scintillator bars and calorimeter blocks were used at the back for the trigger. Photosensors were protected from radiation damage with lead blocks.}
\end{figure}

The Cherenkov telescope was installed in Hall C at Jefferson Lab, as shown in Figure~\ref{fig3}. It was mounted at beamline level pointing towards the target cell and positioned 4.8~m away from the target scattering chamber. Its entrance window was facing the target at an angle of $68^\circ$ with respect to the beam direction. The distance and angle were constrained by space and safety considerations due an ongoing experiment in the hall, resulting in a moderate background-rate environment. Lead blocks shielded the photosensors from direct radiation damage.

The external trigger was provided by scintillator bars and the calorimeter blocks, which were placed behind the exit window of the Cherenkov telescope. Four calorimeter blocks, previously used for the HERA-B calorimeter outer section cells \cite{25}, were stacked together to ensure the full coverage of the exit window area. For trigger purposes, each calorimeter block had one readout channel. The calorimeter threshold was set to $\sim$ 500 mV. This relatively high threshold, far above the minimum-ionizing particle response, reduces the pion background while keeping the trigger rate below the limit of the data acquisition system. The trigger and signals from the scintillator, calorimeter, and photosensors were all fanned-out and sent to two custom-made Jefferson Lab F250 flash Analog-to-Digital Converters (FADC250) \cite{26} for data acquisition. 
During the test beam exercise, we used an 8.1 GeV electron beam, impinging a liquid $H_2$ target. When a high energy particle (mainly scattered electron) traveled through the Cherenkov telescope, filled with CO$_2$ at atmospheric pressure, a cone of Cherenkov photons was emitted along the particle path. The Cherenkov photons were then reflected onto the photosensors by the flat mirror. Simultaneously, the particle continued traveling through the mirror and generated trigger signals in the scintillator and calorimeter, initiating the data acquisition.

\section{Results and Discussion}
\label{sec:Results}
\subsection{Electronics rate estimation}

\begin{figure}[htbp]
\centering 
\includegraphics[width=0.75\textwidth,trim=0 0 0 0,clip]{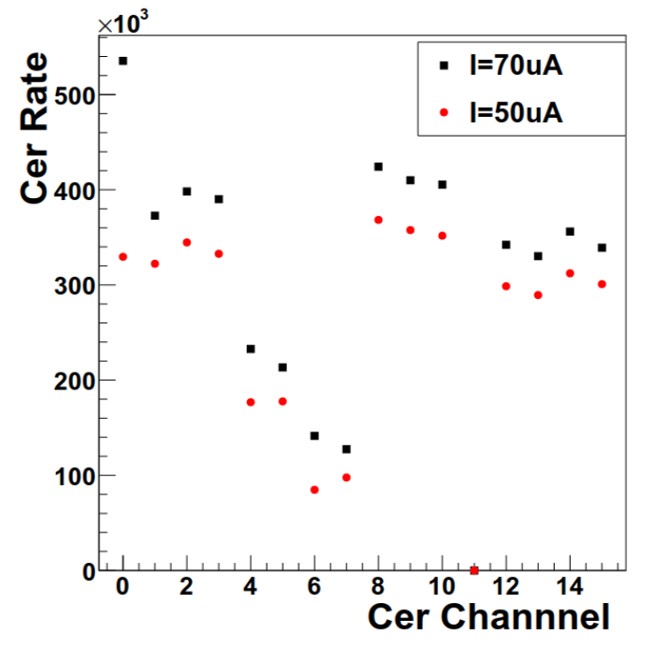}
\caption{\label{fig4} Estimation of the electronics rate for each channel. The black squares correspond to a 70 $\mu A$ beam intensity, while the red circles correspond to a lower 50 $\mu A$ beam intensity.}
\end{figure}

In our experimental conditions, each channel's electronics rate was estimated to be about 400 kHz, which is $\sim1.5$~MHz per MaPMT, or 60~kHz/cm$^2$. This rate is dominated by the single-photon background. Figure~\ref{fig4} shows the estimated electronics rate for each channel. The actual rate differs between the quadrants due to the alignment between the detector, the target and the coarse mirror alignment. 

\subsection{Hamamatsu MaPMT array response}

\begin{figure}[htbp]
\centering 
\includegraphics[width=1\textwidth,trim=0 0 0 0,clip]{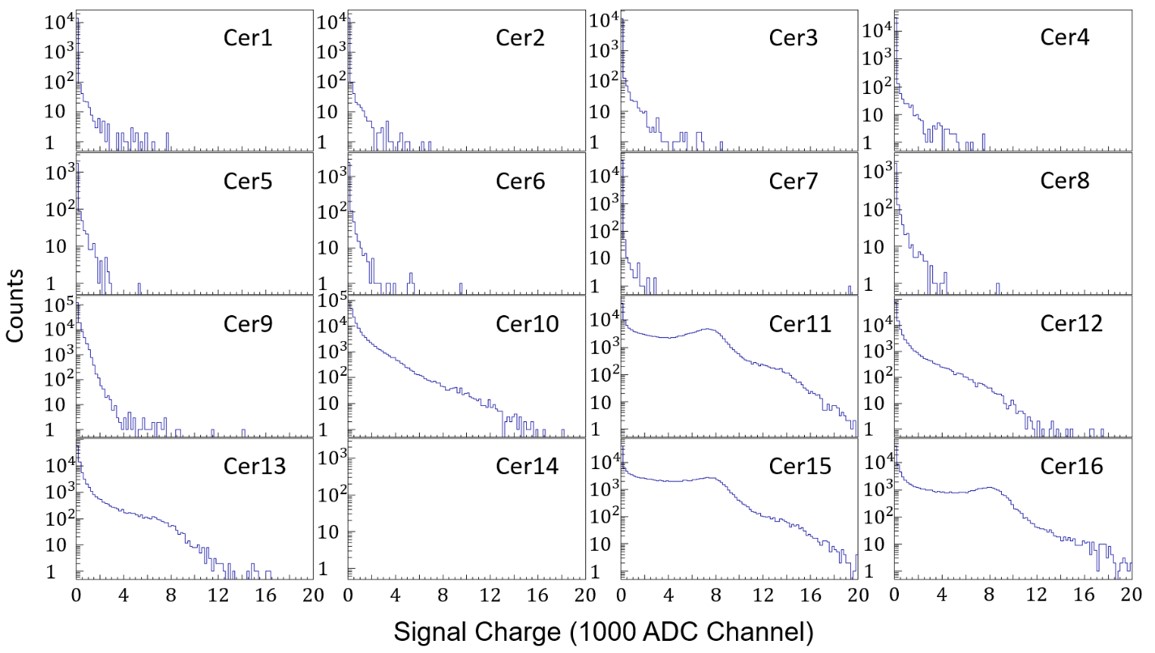}
\caption{\label{fig5} Integrated signal charge distribution of each channel within a $3\sigma$ timing cut. Channel “Cer14” data was not recorded due to a wire connection issue. Notice the high event rate at a small amplitude in each channel due to the high background rate. }
\end{figure}

The background rejection for the test data was mostly accomplished by a timing cut on the difference between the LAPPD signal and the calorimeter signal.
The timing difference of each channel can be well described by a Gaussian distribution with $\sigma < 2$ ns. Figure~\ref{fig5} shows the distribution of the signal amplitude of each channel after a $3\sigma$ timing difference cut. Channel “Cer14” had no data recorded because of a wire connection issue. It can be seen that a significant single photoelectron background dominates the raw MaPMT signals. Signal correlation analysis is necessary to select good events from the background. Meanwhile, signals with large amplitudes were concentrated on channel $8 \sim 15$, indicating that the Cherenkov telescope was not optimally aligned. The Cherenkov photon cone appears shifted to the lower side of the MaPMT array.

\begin{figure}[htbp]
\centering 
\includegraphics[width=1\textwidth,trim=0 0 0 0,clip]{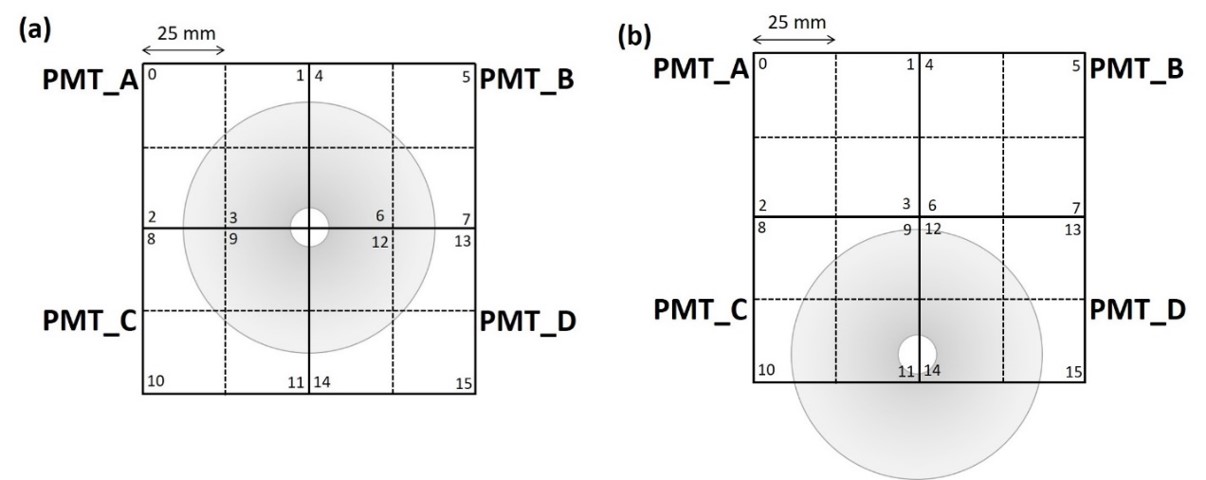}
\caption{\label{fig6} The electronic configuration of the $2 \times 2$ Hamamatsu MaPMT array in the Cherenkov telescope. The shaded area indicates the Cherenkov photon cone in (a) a perfect alignment condition, and (b) identified position shift due to misalignment. In this scenario Cherenkov photons are collected by PMT C and D only.}
\end{figure}

With a perfect alignment to the line-of-sight of the target center, the Cherenkov photon cone generated by an electron along the axis of the tank is expected to cover the central area of the sensors with an outer diameter of 7.7 cm and an inner diameter of 1.2 cm, indicated by the shadow area in Figure~\ref{fig6}(a). Figure~\ref{fig6}(b) illustrated a possible Cherenkov cone positional shift to demonstrate how the misalignment affected the Cherenkov photon collection. Given that the alignment adjustment was not possible due to limited and opportunistic access to the experimental hall, this issue could not be addressed at the time of the measurement. A future experiment with a broader area photosensor coverage and a more suitable alignment, which will mitigate this issue, is discussed in section~\ref{sec:Summary}.

\begin{figure}[htbp]
\centering 
\includegraphics[width=1\textwidth,trim=0 0 0 0,clip]{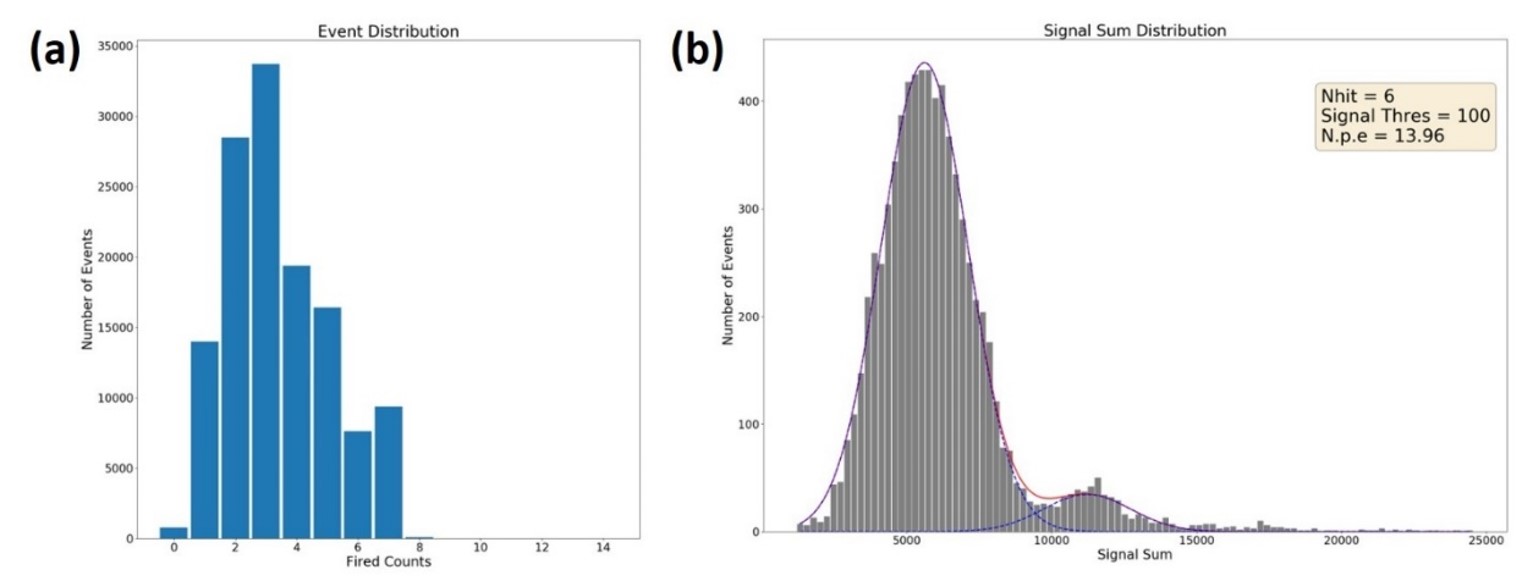}
\caption{\label{fig7} (a) Distribution of the number of fired channels. A fired channel is defined by requiring a signal greater than 100 ADC value after the removal of crosstalk signals. (b) ADC spectra of the integrated signal for the events with 6 fired channels. The two prominent peaks of the spectra are Cherenkov events and pair-production events, respectively.}
\end{figure}

Pixel correlation and high threshold selection were applied in our event-by-event data analysis to reduce the background events. We extracted signal waveforms for all channels from the raw data. After removing crosstalk signals, the distribution of the number of fired channels with a signal threshold at 100 ADC units was obtained and shown in Figure~\ref{fig7}(a). The number of fired channels for each event ranges from 0 to 8. A low fired channel count indicates incomplete acceptance for the Cherenkov cone. Events with more than 5 fired channels were selected for further analysis. Figure~\ref{fig7}(b) shows the ADC spectra of the Cherenkov signal summed over all the channels for the events with 6 fired channels. Two prominent peaks can be identified in the result, with the smaller peak at about twice the ADC value of the higher peak. The latter represents the single electron Cherenkov events, while the former corresponds to the possible conversion of high energy photons to electron-positron pairs originating at the aluminum telescope entrance window. A two-gaussian fit was performed to extract the mean ($\mu$) and standard deviation ($\sigma$) of the peaks, and the mean number of detected photoelectrons ($N_\text{pe}$) was estimated following the equation below \cite{27}:
\[
N_\text{pe} \approx \frac{\mu^2}{\sigma^2}
\]

The calculation gave an experimental $N_\text{pe}$ of 14, half of what was expected from the calculation ($N_\text{pe} \sim$ 30). This large difference was later understood and found to be due to misalignment effects, including mirror misalignment and overal detector pointing misalignment. Misalignment issues are discussed in the GEANT4 simulation section below. They were found to be the root cause of the lowered photodetection acceptance of the MaPMT photosensor array, on average capturing about 50\% of the incident Cherenkov cone.

\subsection{GEANT4 Simulation}

\begin{figure}[htbp]
\centering 
\includegraphics[width=1\textwidth,trim=0 0 0 0,clip]{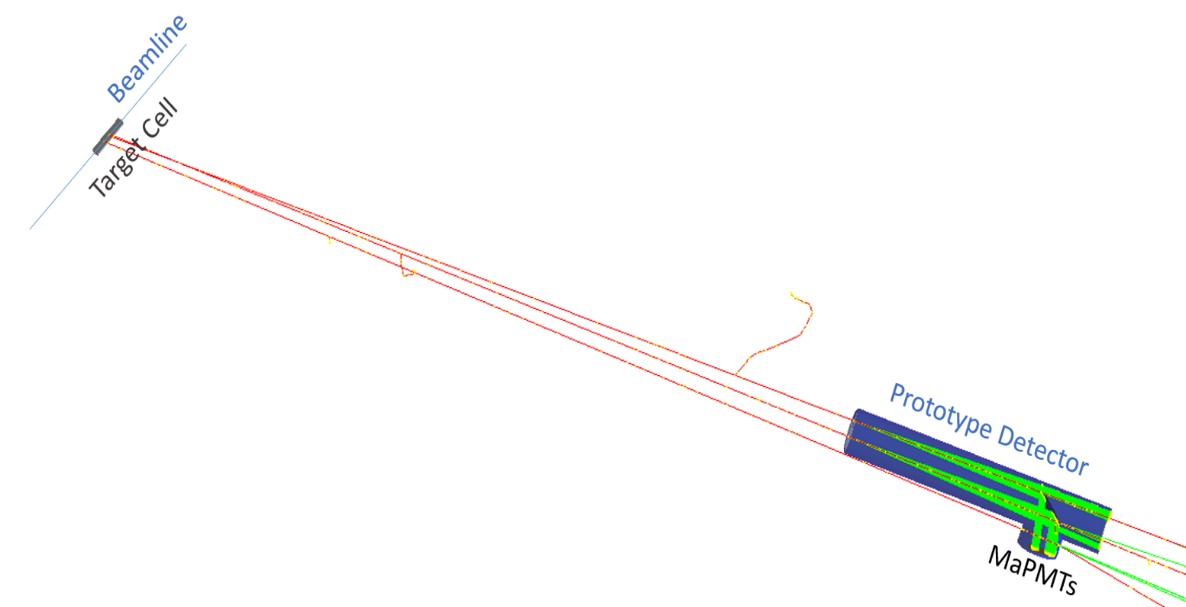}
\caption{\label{fig8} Geometrical setup and the visualization of a few events in the simulation. The red traces represent the incoming electrons, and green traces show the optical photons generated via the Cherenkov process in the CO2 gas. Most of the optical photons are reflected by the mirror and are detected by the MaPMT array.}
\end{figure}

\begin{figure}[htbp]
\centering 
\includegraphics[width=1\textwidth,trim=0 0 0 0,clip]{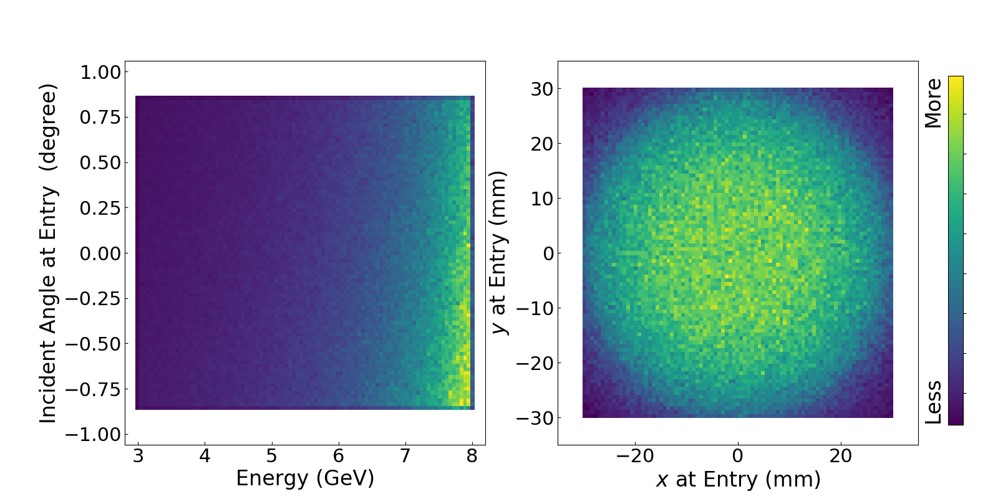}
\caption{\label{fig9} The phase space of incident events that triggered the detector. Angles and positions are with respect to the geometric center of the longitudinal cylindrical tank.}
\end{figure}

\begin{figure}[htbp]
\centering 
\includegraphics[width=1\textwidth,trim=0 0 0 0,clip]{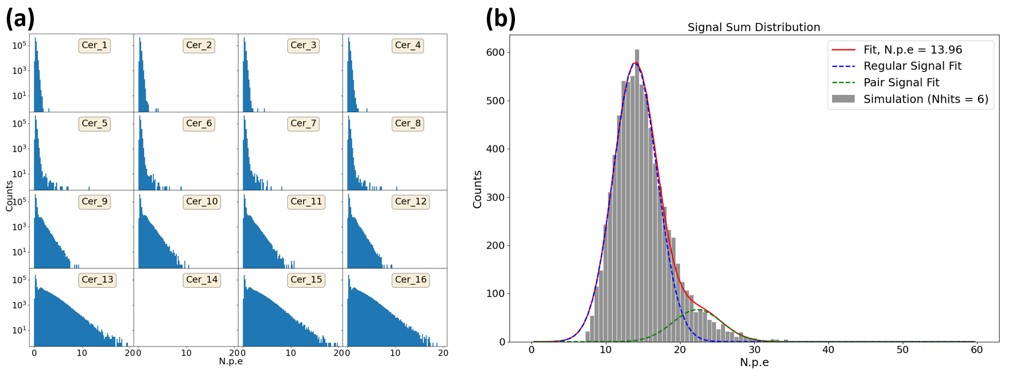}
\caption{\label{fig10} Simulation results with an additional $15^\circ$ tilt of the mirror. (a) Signal amplitude distribution of each channel from the simulation. One channel is turned off to be consistent with the experimental data (see Figure~\ref{fig5}). (b) Sum of signals for the events with 6 fired channels. }
\end{figure}

A detailed simulation was developed with the GEANT4 simulation toolkit  \cite{28} to explain the observed experimental results. 
It simulates the Cherenkov photons' generation process and path for high-energy particles from the target, such as electrons and photons, traveling through the Cherenkov detector. The refractive index of the radiator, the experimentally determined mirror reflectivity, and the quantum efficiency (QE) of the MaPMTs were incorporated in the simulation.
The QE curve used in the simulation also accounts for the effect of the wavelength shifter coating. Figure~\ref{fig8} shows event visualizations from the simulation, which include the geometrical setup starting from the target location, the entrance aluminum window of the detector, the radiator volume and the photosensor array of 4 MaPMTs. In this visualization, incoming electrons (red traces) generate multiple optical photons (green traces) through the Cherenkov process in the CO$_2$ gas. Most of the optical photons are reflected by the mirror onto the photosensor surface. 
In the simulation, electrons and photons are generated at the liquid hydrogen target cell at energies from 3 to 8 GeV. The angular distribution of the incident electrons follows the cross-section of electron scattering off a structureless proton, namely the Mott cross section. In the simulation, particles that deposit more than 0.5 GeV in the calorimeter are accepted. As a result, the subtended phase space of these incident particles is shown in Figure~\ref{fig9}. 
It was found that  an additional mirror tilt angle of $15^\circ$ results in a good agreement with the experimental data, as shown in Figure~\ref{fig10}.
This observed misalignment is most likely caused by a slight pointing misalignment of the detector with respect to the target cell.
A precise survey of this misalignment was not possible due to the parasitic nature of this experiment and radiation concerns.

\subsection{LAPPD with stripline readout}

\begin{figure}[htbp]
\centering 
\includegraphics[width=1\textwidth,trim=0 0 0 0,clip]{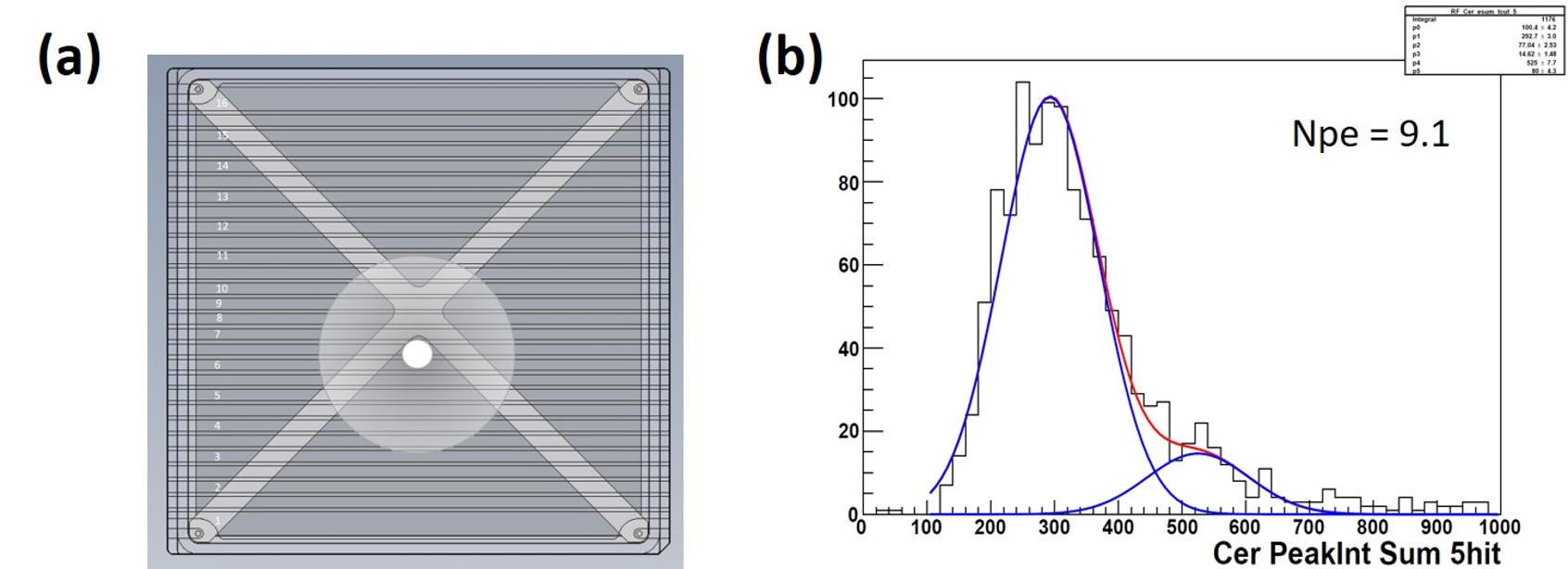}

\caption{\label{fig11} (a) Stripline readout and electronic channel configuration of the LAPPD in the small Cherenkov telescope. The "X" indicates the position of the X cross spacer in the device. (b) ADC spectra of the integrated signal for the events with 5 fired channels.}
\end{figure}

The $20 \times 20$ ~cm$^2$ LAPPD was accommodated in a dark box attached to the Cherenkov tank for photon detection. The stripline readout and electronic channel configuration are shown in Figure~\ref{fig11}(a). The Cherenkov photon cone projection on the LAPPD, with a misalignment similar to that for the MaPMT test, is indicated by the circular shaded area. In this case, however, the large LAPPD active area allowed the collection of all generated Cherenkov photons even with the misalignement effects. 

The same event-by-event data analysis method, including $3\sigma$ timing cut, pixel correlation, and high threshold selection, was applied to the LAPPD data. Figure~\ref{fig11}(b) shows the ADC spectra of the Cherenkov signal integral sum for events where 5 channels fired. The prominent single events and the pair production events are separated in the histogram. The center value of the pair production events is precisely twice that of the single events. The experimental number of photoelectrons $N_\text{pe} = 9.1$ was obtained by fitting the histograms with two Gaussian distributions. The experimental $N_\text{pe}$ value agrees with the expected $N_\text{pe}$ of 10.8 from the calculation for this LAPPD used in our setup. The small $N_\text{pe}$ of the LAPPD setup is expected due to the low QE $\sim 7.3\%$ of this LAPPD, which is 4 times lower than that of the MaPMTs. However, the $N_\text{pe}$ for the LAPPD setup is not 4 times lower than that of the MaPMT setup ($N_\text{pe}$ = 14), due to the full coverage of the Cherenkov cone in the LAPPD setup, compared to the partial coverage of the Cherenkov cone in the MaPMT setup. 

\begin{figure}[htbp]
\centering 
\includegraphics[width=0.75\textwidth,trim=0 0 0 0,clip]{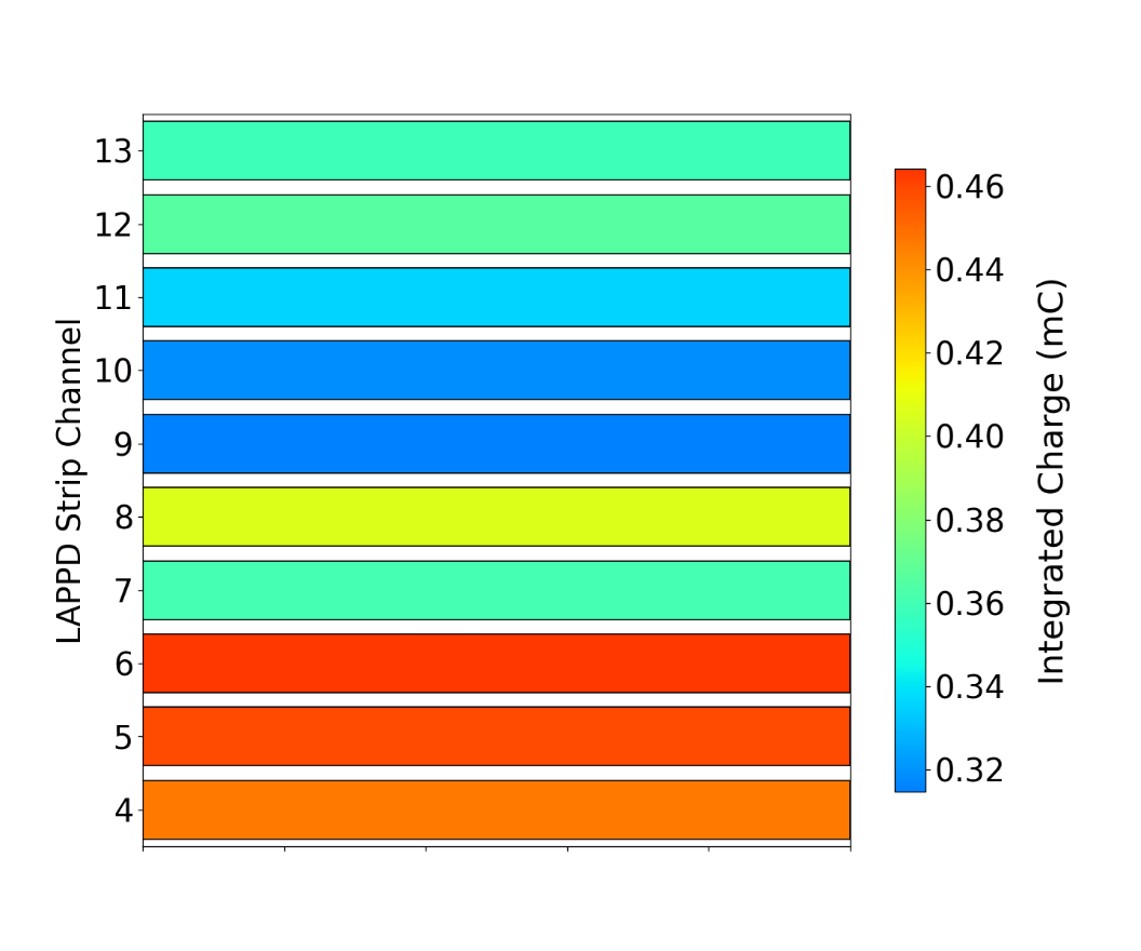}

\caption{\label{fig12} LAPPD occupancy for one experimental run from strip 4 to strip 13, corresponding to the MaPMT geometrical coverage. The bottom strips show higher occupancies, similar to the MaPMT results.}
\end{figure}

The LAPPD results were further analyzed and compared to the MaPMT results. The area of LAPPD sensor is $20 \times 20$ ~cm$^2$, covered by 16 strip readout units which are aligned as shown in Figure~\ref{fig11}(a). The MaPMT array size is $10 \times 10$ ~cm$^2$, covering the area between strip 4 and strip 13 of the LAPPD sensor. Figure~\ref{fig12} shows the LAPPD occupancy (of one experimental run) from strip 4 to strip 13, which corresponds to the MaPMT geometrical coverage. The LAPPD also observes a higher occupancy at the bottom of this area, consistent with the data from MaPMT array. 
This demonstrates that the shift in observed Cherenkov cone position is caused by geometrical effects such as pointing misalignment.

When comparing the ADC spectra using the LAPPD as photosensor to that using the MaPMT array, one notes that the LAPPD exhibits a worse separation of the single events to those of pair-production. This can be traced to the lower QE of this LAPPD. With high-resolution pixelized readout and higher QE, currently under development, we expect LAPPDs to perform significantly better. Furthermore, the spatial information of the signals can then be utilized to perform precise pattern recognition, a feature important for Ring Imaging Cherenkov detectors.

\section{Summary and future work}
\label{sec:Summary}

We reported on first-time beam studies of the performance of commerially available sensors, MaPMTs and and a low-cost internal stripline LAPPD, in an open high-rate/high-background environment at Jefferson Lab using a telescopic Cherenkov device.
Both MaPMTs and the LAPPD were able to operate in such an environment, detect the Cherenkov signals and separate single-electron events from pair-production events.
Despite a slight pointing misalignment of the detector, which reduced the Cherenkov cone acceptance for the MaPMT array by 50\% relative to the LAPPD, a better separation between single-electron and pair-production events was achieved for the MaPMT array due to their higher QE.
The LAPPD performance is expected to improve with high-resolution pixelized readout and higher QE, currently under development.
Our beam studies confirm that the MaPMTs array, internal stripline LAPPD, and the FADC electronics can handle the estimated rate of 60 kHz/cm$^2$, a first step towards validation of these detectors for high-rate/high-background experiments such as the SoLID experimental program at Jefferson Lab.

In light of our results, a beam study with the extreme luminosity conditions expected in the SoLID parity-violation deep-inelastic scattering experiment~\cite{29}, using a Cherenkov prototype with a larger-scale MaPMT array and pixelated LAPPD will be a natural next step. 
This larger-scale Cherenkov setup should be more resilient against pointing misalignment, in order to maximize its Cherenkov photon collection efficiency. Higher QE, pixelated LAPPDs with capacitively coupled readout, under development at Incom, could be a possible advanced photosensor for the SoLID Cherenkov counters. As LAPPD technology continues to evolve rapidly, current sensor performance has already improved considerably since the production of the LAPPD used in our studies.

\end{document}